\documentclass[
 twocolumn,
superscriptaddress,
 amsmath,amssymb,
 aps, 
prb,
]{revtex4-2}
\usepackage{graphicx}
\usepackage{dcolumn}
\usepackage{bm}
\usepackage{mathtools}
\usepackage{color}
\usepackage{appendix}
\usepackage[colorlinks,bookmarks=false,citecolor=blue,linkcolor=black,urlcolor=blue]{hyperref}
\usepackage[nameinlink]{cleveref}
\Crefname{equation}{Equation}{Equations}
\Crefname{figure}{Figure}{Figures}
\Crefname{section}{Section}{Sections}
\Crefname{appendix}{Appendix}{Appendices} 
\Crefname{tabular}{Tabular}{Tabulars}
\crefname{equation}{Eq.}{Eqs.}
\crefname{figure}{Fig.}{Figs.}
\crefname{section}{Sec.}{Secs.}
\crefname{appendix}{App.}{Apps.}
\crefname{tabular}{Tab.}{Tabs.}
\usepackage{tikz}
\usetikzlibrary{external,positioning,shapes.geometric}
\usepackage{pgfplots}
\usepackage{SIunits}
\pgfplotsset{width=7cm,compat=1.18}
\usepackage{pgfplotstable}
\usepgfplotslibrary{groupplots}
\usepackage{placeins}
\usepackage{float}
\usepackage{lipsum}
\usepackage{natbib}

\newcommand{\opensquareline}{\tikz{\node[rectangle,scale=0.7, draw=black!60, fill=white!5,  thick]{};\draw[-] (-0.4,0) to (0.4,0) ; }}
\newcommand{\opencircleline}{\tikz{\node[circle,scale=0.7, draw=black!60, fill=white!5,  thick]{};\draw[-] (-0.4,0) to (0.4,0) ; }}

\makeatletter
\usepackage{etoolbox}
\patchcmd\H@refstepcounter{\protected@edef}{\protected@xdef}{}{}
\makeatother

\begin{document}
\title{Mapping reservoir-enhanced superconductivity to near-long-range magnetic order in the undoped {one-dimensional Anderson and Kondo lattices}}

\author{J. E. Ebot}
\affiliation{SUPA, Institute of Photonics and Quantum Sciences, Heriot-Watt University,
Edinburgh EH14 4AS, United Kingdom}

\author{Lorenzo Pizzino}
\affiliation{DQMP, University of Geneva, 24 Quai Ernest-Ansermet, 1211 Geneva, Switzerland}

\author{Sam Mardazad}
\affiliation{SUPA, Institute of Photonics and Quantum Sciences, Heriot-Watt University,
Edinburgh EH14 4AS, United Kingdom}

\author{Johannes S. Hofmann}
\affiliation{Department of Condensed Matter Physics, Weizmann Institute of Science, Rehovot 76100, Israel}
\affiliation{Max-Planck-Institut f\"ur Physik komplexer Systeme,N\"othnitzer Strasse 38, 01187 Dresden, Germany}

\author{Thierry Giamarchi}
\affiliation{DQMP, University of Geneva, 24 Quai Ernest-Ansermet, 1211 Geneva, Switzerland}

\author{Adrian Kantian}
\affiliation{SUPA, Institute of Photonics and Quantum Sciences, Heriot-Watt University,
Edinburgh EH14 4AS, United Kingdom}
\date{\today}

\begin{abstract}
    The undoped Kondo necklace in 1D is a paradigmatic and well understood model of a Kondo insulator.
    This work performs the first large-scale study of the 1D Anderson-lattice underlying the Kondo necklace with quasi-exact numerical methods, comparing this with the perturbative effective 1D Kondo-necklace model derived from the former.
    This study is based on an exact mapping of the Anderson model to one of a superconducting pairing layer connected to a metallic reservoir which is valid in arbitrary spatial dimensions, thereby linking the previously disparate areas of 
    reservoir-enhanced superconductivity, following Kivelson's pioneering proposals, and that of periodic Kondo-systems. 
    Our work reveals that below the length-scales on which the insulating state sets in, which can be very large, superconducting and density-density correlations are degenerate and may both appear to approach an almost ordered state, to a degree that far exceeds that of any isolated 1D pairing layer with short-range interactions.
    We trace these effects to the effective extended-range coupling that the metallic layer mediates within the pairing layer.
    These results translate directly to the appearance of near-long-range magnetic order at intermediate scales in the Kondo-systems, and explain the strong renormalization of the RKKY-coupling that we effectively observe, in terms of the back-action of the pairing layer onto the metallic layer.
    The effects we predict could be tested either by local probes of quasi-1D heavy fermion compounds such as CeCo$_2$Ga$_8$, in engineered chains of ad-atoms or in ultracold atomic gases.
\end{abstract}

\maketitle
\section{Introduction}\label{sec:intro}
The phase coherence of pairs and the pair-field amplitude of a superconducting state are two antagonistic quantum properties.
Both are essential to realise superconducting off-diagonal long range order (ODLRO), or, due to the Mermin-Wagner theorem (MWT)~\cite{Thierrybook2003}, algebraically decaying quasi-order in the best-case outcome for low-dimensional systems with short-range couplings and interactions.
For both conventionally mediated as well as for high-temperature superconductors (HTS), there is substantial evidence that an optimal $T_{\rm c}$ is obtained around the region where the amplitude- and  phase-dominated regimes cross over into one another~\cite{Emery1995,Fontenele2022}.
This has lead to Kivelson's famous proposal on boosting superconductivity using bilayer setups, maximising the superconducting pair-field in one layer while simultaneously maximising the superconducting stiffness by use of the second layer~\cite{KIVELSON2002}. 

This prospect, of attaining superconducting properties superior to those of the isolated pairing layer by sidestepping the competition between pair-phase and pair-amplitude that had previously been seen as unavoidable, has triggered a number of experimental and theoretical studies~\cite{Yuli2008,Erez2008,Lobos2009,Parker2010,Gideon2012,JEEBOT2025}.
Recently, our own work on the 1D version of Kivelson's proposal using quasi-exact numerical methods has shown that a 1D metal with weak tunnel-coupling to a layer with intrinsic pairing can lead to far better superconducting properties than the isolated pairing layer~\cite{JEEBOT2025}.
This is due to ranged pair-pair coupling mediated by the metal, which greatly boosts superconducting stiffness, while the potentially negative impact of a depressed pair-amplitude due to the reverse proximity effect of the metal on the pairing layer can be abated by detuning the nominal Fermi points of the two layers away from each other.
While the proximity effect emanating from the pairing layer into the metallic layer ultimately limits the range of the mediated pair-pair coupling, superconducting correlations may still decay algebraically with such low exponents as to closely approach ODLRO~\cite{JEEBOT2025}.
This rich and complex physics emerges already for a basic model system of Kivelson's proposal, such as the one we studied previously.
Crucially, much of this complexity stems from the back-actions of the two different layers onto another that we were able to capture with quasi-exact numerics, but in order to get analytical insight into how these properties of the layers may help or hinder superconducting performance, an analytical framework remains to be developed, which needs to go beyond previous work that had not taken back-actions into account~\cite{Erez2008,Lobos2009,Gideon2012}

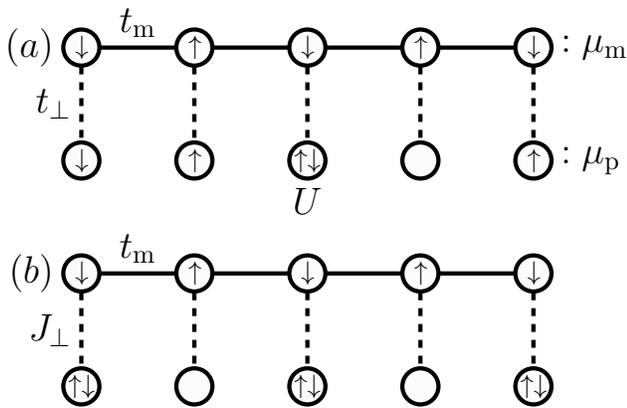
\begin{figure}[t!]
  \centering
\hspace*{0cm}\begin{tikzpicture}
    [
     dotnode/.style={circle, draw=black!100, fill=black!1, ultra thick, minimum size=12mm,text centered,text width=0.2cm,align=center,scale=0.4},
    circlenode/.style={circle, draw=black!100, fill=black!1, ultra thick, minimum size=12mm,text centered,text width=0.2cm,align=center,scale=0.4},
	]
 
        \node[dotnode]      (dotsc1)  {\Huge{$\hspace*{-0.1cm}\downarrow$}};
        \node[dotnode]      (dotsc2)  [left=of dotsc1] {\Huge{$\hspace*{-0.1cm}\uparrow$}};
        \node[dotnode]      (dotsc3)  [left=of dotsc2] {\Huge{$\hspace*{-0.1cm}\downarrow$}};
        \node[dotnode]      (dotsc4)  [left=of dotsc3] {\Huge{$\hspace*{-0.1cm}\uparrow$}};
        \node[dotnode]      (dotsc5)  [left=of dotsc4] {\Huge{$\hspace*{-0.1cm}\downarrow$}};
        \node [black,right] at (dotsc1.east) {\Large{$: \mu _{\rm m}$}};
        \node [black,left] at (dotsc5.west) {\hspace*{-0.0cm}{\Large{$(a) $}}};
        \node[dotnode]      (dotme1)  [below=of dotsc1] {\Huge{$\hspace*{-0.1cm}\uparrow$}};
        \node[dotnode]      (dotme2)  [left=of dotme1] {};
        \node[dotnode]      (dotme3)  [left=of dotme2] {\Huge{$\hspace*{-0.33cm}\uparrow\downarrow$}};
        \node[dotnode]      (dotme4)  [left=of dotme3] {\Huge{$\hspace*{-0.1cm}\uparrow$}};
        \node[dotnode]      (dotme5)  [left=of dotme4] {\Huge{$\hspace*{-0.1cm}\downarrow$}};
        
        \node [black,right] at (dotme1.east) {\Large{$: \mu _{\rm p}$}};
        \node [black,below] at (dotme3.south) {\Large{$U$}};
        
        \draw[dashed,ultra thick] (dotsc1.south)  to  (dotme1.north) ;
        \draw[dashed,ultra thick] (dotsc2.south)  to  (dotme2.north) ;
        \draw[dashed,ultra thick] (dotsc3.south)  to  (dotme3.north) ;
        \draw[dashed,ultra thick] (dotsc4.south)  to  (dotme4.north) ;
        \draw[dashed,ultra thick] (dotsc5.south)  to node[anchor=east]{\Large{$t_\perp $} } (dotme5.north) ;
        \draw[-,ultra thick] (dotsc2.east)  to  (dotsc1.west) ;
        \draw[-,ultra thick] (dotsc3.east)  to  (dotsc2.west) ;
        \draw[-,ultra thick] (dotsc4.east)  to  (dotsc3.west) ;
        \draw[-,ultra thick] (dotsc5.east)  to node[anchor=south]{\Large{$t_{\rm m}$} } (dotsc4.west) ;
        \node[circlenode]      (sc1)  [below=of dotme1] {\Huge{$\hspace*{-0.1cm}\downarrow$}};
        \node[circlenode]      (sc2)  [left=of sc1] {\Huge{$\hspace*{-0.1cm}\uparrow$}};
        \node[circlenode]      (sc3)  [left=of sc2] {\Huge{$\hspace*{-0.1cm}\downarrow$}};
        \node[circlenode]      (sc4)  [left=of sc3] {\Huge{$\hspace*{-0.1cm}\uparrow$}};
        \node[circlenode]      (sc5)  [left=of sc4] {\Huge{$\hspace*{-0.1cm}\downarrow$}};
        \node [black,left] at (sc5.west) {\hspace*{0cm}{\Large{$(b) $}}};
        \node[circlenode]      (m1)  [below=of sc1] {\Huge{$\hspace*{-0.33cm}\uparrow \downarrow$}};
        \node[circlenode]      (m2)  [left=of m1] {};
        \node[circlenode]      (m3)  [left=of m2] {\Huge{$\hspace*{-0.33cm}\uparrow \downarrow$}};
        \node[circlenode]      (m4)  [left=of m3] {};
        \node[circlenode]      (m5)  [left=of m4] {\Huge{$\hspace*{-0.33cm}\uparrow \downarrow$}};
        \draw[dashed,ultra thick] (sc1.south)  to  (m1.north) ;
        \draw[dashed,ultra thick] (sc2.south)  to  (m2.north) ;
        \draw[dashed,ultra thick] (sc3.south)  to  (m3.north) ;
        \draw[dashed,ultra thick] (sc4.south)  to  (m4.north) ;
        \draw[dashed,ultra thick] (sc5.south)  to node[anchor=east]{\Large{$J_\perp $} } (m5.north) ;
        \draw[-,ultra thick] (sc2.east)  to  (sc1.west) ;
        \draw[-,ultra thick] (sc3.east)  to  (sc2.west) ;
        \draw[-,ultra thick] (sc4.east)  to  (sc3.west) ;
        \draw[-,ultra thick] (sc5.east)  to node[anchor=south]{\Large{$t_{\rm m}$} } (sc4.west) ;
\end{tikzpicture}
\caption{\textbf{(a)} Schematic figure of the 1D Kivelson bilayer proposal, $\hat{H}_{\rm KBP}$ at half-filling, \cref{model_equ}).
\textbf{(b)} Kondo necklace: effective model \cref{eq:hkbpj} resulting from \cref{model_equ} using Schrieffer-Wolff transformation.
}
\label{fig:Model_system}
\end{figure}

A first start for such a framework is to further simplify the set-up, where the pairing layer consists just of negative-$U$ centres, the pairing sites, with direct tunneling between these sites set to zero, as shown in~\cref{fig:Model_system}.
As detailed in the following sections, such a set-up allows us to use an exact mapping to make contact with the known results for another group of much-studied hybrid bilayer systems, which superficially appear to be quite different, the 1D Anderson-lattices~\cite{Guerrero1996,Guerrero2001,Luo2002,Batista2003,Wang2006,Yu2008,Bertussi2011,Gulasci2015} and Kondo-lattices~\cite{Jullien1981,Sigrist1992,Yu1993,Lavagna1998,Shibata1999,BookSectionShibata1999,McCulloch2002,Schauerte2005}.
Where the Kondo-effect is driven by isolated localized electronic  impurities in a metal~\cite{Bickers1987}, the undoped, i.e. half-filled, Anderson-lattice consists of a regular lattice of localized orbitals each occupied by a single electron, which are coupled to itinerant conduction-band electrons.
The Kondo-lattice is then derived from the Anderson-lattice in the perturbative regime, where Coulomb-repulsion in the localized orbitals is large compared to the coupling between those orbitals and the conduction band, and relative chemical potentials between the layers are adjusted with respect to Coulomb-repulsion such as to keep the electron density of the localised orbitals fixed at one~\cite{Gulasci2015}.
These models arose in the context of the heavy-electron materials, in which the lattice of localized spins is realized by $4f$- or $5f$-ions such as Ce or U respectively~\cite{Stewart1984,Das2014,Wang2017,Fumega2024}.
The doped and undoped 1D Kondo-lattice, termed Kondo-necklace, has been intensely studied since Doniach's initial investigations~\cite{Doniach1977}.
Undoped, the Kondo-necklace has been established to be gapped for arbitrary coupling strengths not just in the charge sector, but also in the spin sector~\cite{Jullien1981,Clare1993}.
In contrast, in 2D the Kondo-lattice appears to be able to close the spin-gap at weaker coupling, realising an antiferromagnetic (AFM) phase in the process, and exhibiting a second-order transition to a fully gapped insulator at stronger coupling~\cite{Assaad1999,Eder2016}.
In comparison to the Kondo-necklace, the underlying 1D Anderson-lattice has been studied significantly less~, and even in the undoped state results on the physics realized at low or zero temperature from reliable methods are lacking either for large finite systems or in the thermodynamic limit\cite{Guerrero1996,Wang2006,Bertussi2011}.

%
%
In the present work, we exploit an exact particle-hole mapping that provides a previously unrealized connection between the very simplest 1D models of Kivelson's proposal for boosting superconductivity at unit electron density - a line of pairing sites connected to a non-interacting 1D metal - and the undoped 1D Anderson- and Kondo-lattices.
Combining quasi-exact numerics at both zero and finite temperature~\cite{Schollwock2011,10.21468/SciPostPhysCodeb.1-v2.4,10.21468/SciPostPhysCodeb.1-r2.4} for large systems with an exact particle-hole transformation, we thus obtain novel insights and results for both these classes of systems.
We demonstrate that these systems appear to approach, respectively, coexisting superconducting and charge-density wave (CDW) long-range order for the 1D Kivelson bilayer systems, and AFM order for the 1D Anderson-/Kondo-lattices, across large scales when coupling to the metal is weak, as the gap that would suppress these orders in the thermodynamic limit becomes very small.
We explain this phenomenon in terms of the ranged pair-pair coupling mediated by the metallic layer, which has been analyzed in various ways for different versions of Kivelson's proposal~\cite{Erez2008,Lobos2009,Gideon2012}, and which is known as RKKY-coupling for the Anderson/Kondo-lattices~\cite{Rusin2017,Nejati2017}.
The present work demonstrates that, in contrast to previous studies that do not account for the back-action of one layer onto the other layer, the coupling mediated by a clean metal is not truly long-ranged, i.e. they do not decay as a low-exponent power law as earlier analysis had indicated, but instead acquire an exponentially decaying envelope.
As would ultimately be expected for any 1D system containing only short-range microscopic couplings, this back-action-driven modification of the metal-mediated coupling prevents these systems from escaping the limitations of the MWT and from actually achieving long-range order.
This work provides the necessary insight for further explorations of these dual systems, now in the doped regime, i.e. away from electronic unit density that leads to these systems being fully gapped, for which we are preparing a companion publication \cite{JEEBOT2025_2}.

This work is structured as follows: 
in~\cref{sec:models}, we introduce the 1D Kivelson bilayer models as well as the dual 1D Anderson-/Kondo-lattices, their respective observables and the mapping between them.
In~\cref{sec:ordering} we show the pair-pair correlation functions we extract from quasi-exact numerics on the 1D Kivelson bilayer models, their change with the strength of the coupling connecting pairing layer and 1D metal, and their interpretation in light of our particle-hole mapping and the known results on the Kondo-necklace as well our own prior work.
In particular, we demonstrate that in the weak-coupling regime these systems appear to closely approach coexisting long-range superconducting and CDW order (or AFM order in the dual systems) on any length scale that can be simulated with the current state of the art, as the gap that suppresses this behaviour in the thermodynamic limit becomes very small.
In this, we combine zero-temperature density matrix renormalization group (DMRG) calculations, as implemented in the SyTen-package~\cite{syten} with those obtained from an auxiliary-field Quantum Monte Carlo (AFQMC) approach, as implemented in the ALF-package~\cite{10.21468/SciPostPhysCodeb.1-v2.4,10.21468/SciPostPhysCodeb.1-r2.4}.
In~\cref{sec:rkky}, we show how the modification of the 1D metal by the back-action of the pairing layer is correlated with the appearance of approaching long-range order at large scales.
We also relate these effects to the RKKY-interaction, the ranged spin-spin interaction mediated via the metallic layer, for the 1D Anderson- and Kondo-lattices.
Finally, in~\cref{sec:disc_and_outl} we discuss the outlook for these systems when doping them, as well as how our predictions may be tested experimentally.
We also discuss the prospects to leverage Kondo-lattice materials to indirectly study Kivelson's bilayer proposal for reservoir-enhanced superconductivity.
\section{Models, Observables \& prior Results}\label{sec:models}
We start from the simplest 1D model of Kivelson's proposal for stabilization of superconductivity in spin-$1/2$ electrons by metallic reservoir.
As shown in~\cref{fig:Model_system}a, this consists of a pairing layer comprised of $L$ isolated sites with an attractive on-site interaction $-U$, $U>0$, connected to a non-interacting 1D metal with tunnel-coupling $t_\perp$, also with $L$ sites.
The metal is modeled via a tight-binding Hamiltonian with tunneling-amplitude $t_{\rm m}$.
The microscopic Hamiltonian of Kivelson's bilayer proposal can thus be written as
\begin{align}
    \hat{H}_{\rm KBP} = \hat{H}_{m} + \hat{H}_{U} + \hat{H}_{\perp},
    \label{model_equ}
\end{align}
where the free electron Hamiltonian is given by 
\begin{align}
        \hat{H}_{m} = \sum_{i,\sigma}^{L} -t_{m}(\hat{c}_{i,m,\sigma}^{\dagger}\hat{c}_{i+1,m,\sigma} + {\rm h.c.})-\mu_{\rm m} \left(\hat{n}_{i,m,\sigma}-\frac{1}{2}\right).
\end{align}
The pairing-sites, and the perpendicular Hamiltonian that couples the free electron and pairing-sites layer, are defined as:
\begin{align}
     \hat{H}_{U} &=  -U\sum_{i}^{L}\hat{n}_{i,p,\uparrow}\hat{n}_{i,p,\downarrow} - \mu_{\rm p}\sum_{i,\sigma}^{L} \left(\hat{n}_{i,p,\sigma}-\frac{1}{2}\right),\label{eq:hu}\\
     \hat{H}_{\perp} &=-{t}_\perp \sum_{i,\sigma}^{L}(\hat{c}_{i,p,\sigma}^{\dagger}\hat{c}_{i,m,\sigma} + {\rm h.c.}),
\end{align}
respectively.
Here, $p$ and $m$ denote the pairing sites and the metallic layer respectively. We define ${\lambda=p,m}$, then $\hat{c}_{i,\lambda,\sigma}$ ($\hat{c}^\dagger _{i,\lambda,\sigma}$) are fermion operators that annihilate (create) an electron with spin $\sigma(=\uparrow,\downarrow)$ at site $i$ in layer $\lambda$, and the density-operators are given by ${\hat{n}_{i,\lambda,\sigma}=\hat{c}^\dagger_{i,\lambda,\sigma}\hat{c}_{i,\lambda,\sigma}}$, and the total density operator is given by ${\hat{n}_{i,\lambda}=\hat{n}_{i,\lambda,\uparrow}+\hat{n}_{i,\lambda,\downarrow}}$.
The chemical potential $\mu_{\lambda}$ adjusts the density of the respective layer.
Throughout this work the density of the respective layers are set to half-filling, $\langle \hat{n}_{i,\lambda,\sigma}\rangle=1/2$.
In the grand-canonical regime, this is achieved by setting  ${\mu_p=-U/2}$, $\mu_m=0$, and in the canonical regime by fixing the total number of electrons to be ${N_c=2L}$, and the total $z$-spin of the system to be $S_z=0$, while ensuring that ${\mu_p-\mu_m=-U/2}$.

In the perturbative regime, where $U\gg t_\perp$, $\hat{H}_{\rm KBP}$ can be approximated as
\begin{equation}
    \begin{aligned}\label{eq:hkbpj}
       \hat{H}^{J_\perp}_{\rm KBP} &= \hat{H}_{m} 
     + J_{\rm z}/4\sum_{i}^{L}(\hat{n}_{i,p} - 1)(\hat{n}_{i,m} - 1) \\
     &-J_{\rm xy}/2 \sum_{i}^{L}(c_{i,p,\uparrow}^{\dagger} c_{i,m,\downarrow}^{\dagger} c_{i,m,\uparrow} c_{i,p,\downarrow} + {\rm h.c.})
    \end{aligned}
\end{equation}
using a Schrieffer-Wolff transformation, where ${J_{\rm xy}=J_{\rm z}=J_{\perp}=\frac{4t_\perp ^{2}}{U}}$, and the possible occupancies on the pairing sites are constrained to either zero or doublons, c.f.~\cref{fig:Model_system}b.

These two Hamiltonians can be mapped exactly to those of the 1D Anderson- and Kondo-lattices respectively.
Using a selective particle-hole transformation on the $\downarrow$-spins
\begin{equation}
    \begin{aligned}\label{eq:phtrafo}
        &\hat{c}^{\dagger}_{i,\lambda,\uparrow} \rightarrow \hat{d}^{\dagger}_{i,\lambda,\uparrow} \\
        &\hat{c}^{\dagger}_{i,p,\downarrow} \rightarrow (-1)^j\hat{d}^{}_{i,p,\downarrow} \quad
        \hat{c}^{\dagger}_{i,m,\downarrow} \rightarrow (-1)^{j+1}\hat{d}^{}_{i,m,\downarrow},\\
    \end{aligned}
\end{equation}
at the half-filling point ${\mu_p=-U/2}$, $\mu_m=0$ we map
\begin{equation}
    \begin{aligned}\label{eq:hal}
        \hat{H}_{\rm KBP}\rightarrow\hat{H}_{\rm AL} &= -t_{\rm m}\sum_{i,\sigma}^L(\hat{d}^\dagger_{i,m,\sigma}\hat{d}^{}_{i+1,m,\sigma}+\text{h.c.})\\
        &+U\sum_i \left ( \hat{n}_{i,p,\uparrow} -\frac{1}{2}\right ) \left ( \hat{n}_{j,p,\downarrow} -\frac{1}{2}\right )\\
        &-t_\perp\sum_{i,\sigma}^L(\hat{d}^{\dagger}_{i,p,\sigma} \hat{d}^{}_{i,m,\sigma} + \text{h.c.})  
    \end{aligned}
\end{equation}
to obtain the Hamiltonian for the 1D Anderson-lattice, $\hat{H}_{\rm AL}$, from $\hat{H}_{\rm KBP}$.
After this mapping, we have ${\hat{n}_{i,\lambda,\sigma}=\hat{d}^\dagger_{i,\lambda,\sigma}\hat{d}_{i,\lambda,\sigma}}$ of course.
Likewise, we obtain the Hamiltonian $\hat{H}_{\rm KL}$ for the 1D Kondo-lattice from $\hat{H}^{J_\perp}_{\rm KBP}$ by the same mapping:
\begin{equation}
    \begin{aligned}\label{eq:hkl}
        \hat{H}^{J_\perp}_{\rm KBP}\rightarrow\hat{H}_{\rm KL} &= -t_{\rm m}\sum_{i,\sigma}^L(\hat{d}^\dagger_{i,m,\sigma}\hat{d}^{}_{i+1,m,\sigma}+\text{h.c.})\\
        &+ J_{\rm z}/4\sum_{i}^{L}(\hat{n}_{i,p} - 1)(\hat{n}_{i,m} - 1) \\
    &+J_{\rm xy}/2 \sum_{i}^{L}(\hat{d}_{i,p,\uparrow}^{\dagger} \hat{d}_{i,m,\downarrow}^{\dagger} \hat{d}_{i,m,\uparrow} \hat{d}_{i,p,\downarrow} + {\rm h.c.})\\
     &= -t_{\rm m}\sum_{i,\sigma}^L(\hat{d}^\dagger_{i,m,\sigma}\hat{d}^{}_{i+1,m,\sigma}+\text{h.c.})\\
        &+ J_{\perp}\sum_{i}^{L}\vec{S}_{i,m} \cdot \vec{S}_{i,p}
    \end{aligned}
\end{equation}
The vector of spin operators ${\vec{S}_{i,\lambda}=(\hat{S}^x_{i,\lambda},\hat{S}^y_{i,\lambda},\hat{S}^z_{i,\lambda})}$, expressed in terms of the underlying fermionic operators, is provided in~\cref{app:degen}, along with the details of the mapping procedure.

In this work, we numerically compute observables for $\hat{H}_{\rm KBP}$ and $\hat{H}^{J_\perp}_{\rm KBP}$, i.e. for the systems with attractive interactions, then transfer the insights gained for these to $\hat{H}_{\rm AL}$ and $\hat{H}_{\rm KL}$.
Likewise, prior results that were obtained for $\hat{H}_{\rm AL}$ and $\hat{H}_{\rm KL}$, i.e. on the repulsively interacting side, are then transferred to $\hat{H}_{\rm KBP}$ and $\hat{H}^{J_\perp}_{\rm KBP}$.
The main observables that we compute numerically are pair-pair, density-density and single-electron correlation functions,  $C_{\lambda}(i,j)$, $N_{\lambda}(i,j)$ and $S_{\rm m}(i,j)$ respectively:
\begin{eqnarray}\label{eq:corrs_1}
    C_{\lambda}(i,j) & = &  \left\langle  \hat{c}^{\dagger}_{i,\uparrow,\lambda} \hat{c}^\dagger_{i,\downarrow,\lambda} \hat{c}_{j,\downarrow,\lambda} \hat{c}_{j,\uparrow,\lambda} \right\rangle ,\\
    \label{eq:corrs_2} S_{\rm m}(i,j) & = &  \left\langle  \hat{c}^{\dagger}_{i,\sigma,\rm m}  \hat{c}_{j,\sigma,\rm m} \right\rangle , \\
    \label{eq:corrs_3} N_{\lambda}(i,j) &=& \left\langle  (\hat{n}_{i,\lambda}-1)(\hat{n}_{j,\lambda}-1) \right\rangle . 
\end{eqnarray}
At $T=0$, we also compute the charge gaps for both Hamiltonians at different $L$-values,
\begin{equation}\label{eq:cgap}
\Delta_c = E_{\rm GS}(2L+2,0) + E_{\rm GS}(2L-2,0) - 2E_{\rm GS}(2L,0),
\end{equation}
where $E_{\rm GS}(N,S_z)$ denotes the ground state energy of a system of length $L$ with $N$ electrons and total $z$-spin of $S_z$.
All of this is done with DMRG as implemented in the SyTen-package~\cite{syten}.
At $T>0$ we also compute the correlation functions~\cref{eq:corrs_1} -~\cref{eq:corrs_3} using AFQMC as implemented in the ALF-package~\cite{10.21468/SciPostPhysCodeb.1-v2.4,10.21468/SciPostPhysCodeb.1-r2.4}. 
Further details on the calculation-parameters are supplied in~\cref{app:sims}.

For $\hat{H}_{\rm KBP}$, we target the regime ${t_{\perp} < \Delta_{\rm p}}$ where $\Delta _{\rm p}$ designates the paring gap of the pairing layer~\cite{Erez2008,Lobos2009,JEEBOT2025}.
As we deliberately omit any kinetic energy terms from the description of the pairing layer, here we have ${\Delta_{\rm p}=U}$.
%

\begin{figure}
    \centering
    \includegraphics[scale=1]{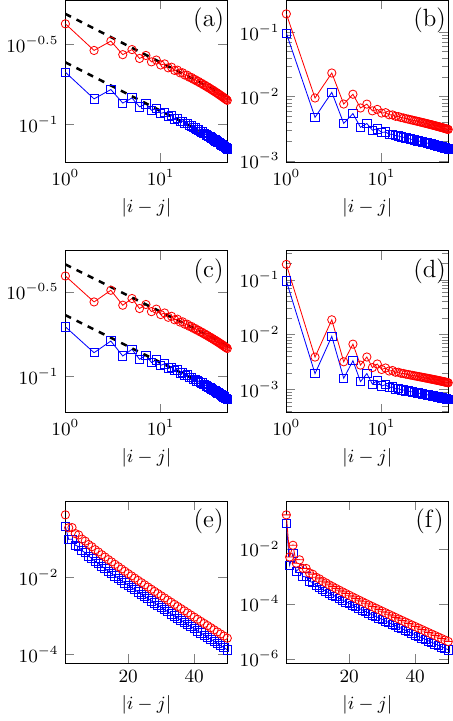}
    \caption{
    $C_{\rm p}(i,j)$ (left column, \protect\opensquareline), $|N_{\rm p}(i,j)|$ (left column, \protect\opencircleline),  $C_{\rm m}(i,j)$ (right column, \protect\opensquareline) and $|N_{\rm m}(i,j)|$ (right column, \protect\opencircleline) with $L=200$.
     \textbf{(a)} $\&$ \textbf{(b)} $U=4, t_{\perp}=0.55,t_{\rm m}=1$  
     \textbf{(c)} $\&$ \textbf{(d)} $U=10, t_{\perp}=3.0,t_{\rm m}=10$ 
     \textbf{(e)} $\&$ \textbf{(f)} $U=10, t_{\perp}=4.0,t_{\rm m}=10.$
    }
    \label{fig 1}
\end{figure}
This work exploits the connections between these four Hamiltonians, via exact particle-hole mapping and second-order perturbation theory, and combines our findings with prior results, leveraging insight gained on one Hamiltonian to the others.
We thus link the previously separate fields of reservoir-enhanced superconductivity and Anderson-/Kondo-lattices.
Prior results for the 1D half-filled Kondo-lattice indicate that the ground state of this Hamiltonian might be fully gapped at any value of ${J_\perp>0}$, i.e. in both the charge- and the spin-sector~\cite{Jullien1981,Yu1993}, thus realising a fully gapped Kondo-insulator.
Numerical results, which had so far been limited to $J\geq t_{\rm m}/2$, indicate that the spin-gap $\Delta_s$ to generally be below the charge gap $\Delta_c$~\cite{Yu1993}, and with ${\Delta_s\ll\Delta_c}$ for ${J_\perp<t_{\rm m}}$.
Early analytical work had indicated that for perturbative values of $J$, the spin-gap might scale as ${\Delta_s\propto e^{-A/J_\perp}}$, with $A$ being a constant that depends on the microscopic physics~\cite{Jullien1981} .
With such a scaling, $\Delta_s$ could easily become too small to be detectable even for the largest finite systems amenable to numerics, and even in experiments, depending on the realization.
Prior to this work, the low-$J_\perp$ regime at half-filling had not been studied systematically and for large systems.
For the underlying Anderson-lattice at half-filling, even fewer previous studies were conducted, limited to ${L\leq 50}$, not systematically addressing the weak-coupling regime, and not investigating the behaviour of the spin-gap.

Under the particle-hole transformation, spin- and charge-gaps are exchanged.
this would suggest a possibility for the charge-gap to scale as ${\Delta_c\propto e^{-A/J_\perp}}$ for the ground state of $\hat{H}_{\rm KBP}^{J_\perp}$ at half-filling.
One of our key results in~\cref{sec:ordering} is that for all our studied parameters the charge-gap of the half-filled 1D Anderson-lattice is well above that of the equivalent 1D Kondo-lattice.
Yet, we find that at sufficiently weak coupling the ground state of $\hat{H}_{\rm KBP}$ exhibits very slowly decaying power-law behaviour for the superconducting pair-pair correlations, in line with our previous work on $\hat{H}_{\rm KBP}$ away from half-filling and with direct non-zero tunneling between the pairing sites~\cite{JEEBOT2025}, on any system sizes that we can simulate.
This behaviour goes along with charge-gaps that extrapolate to values indistinguishable from zero in the limit ${L\rightarrow\infty}$.
In those cases, we therefore fit the pair-pair correlation $C_{\lambda}(i,j) $ with a power law
\begin{equation}\label{eq:cfit}
C_{\lambda}(i,j)=A_{C,\lambda}|i-j|^{K_\lambda^{-1}}.
\end{equation}
In this way, we assign a Tomonaga-Luttinger liquid (TLL) parameter $K_{\lambda}$ to those large but finite systems that appear to exhibit ${\Delta_c=0}$.
As in any generic TLL, a lower value of $K_{\lambda}^{-1}$ thus implies pair-pair correlations that decay more slowly, and as $K_{\lambda}^{-1}$ decreases and appears to approach zero, one can speak of near-long-range-order, as $C_{\lambda}(i,j)$ seems to decay very little across the entirety of the system.
Also in line with~\cite{JEEBOT2025}, we find that the single particle density matrix of the metal, $S_{\rm m}(i,j)$, always decays exponentially at any finite coupling to the pairing layer, as opposed to the long-range algebraic behaviour of the metal at zero coupling.
We will discuss the resulting physics and their implications for the properties of $\hat{H}_{\rm AL}$ and $\hat{H}_{\rm KL}$ in~\cref{sec:rkky}.

\section{Different (apparent) orders at different scales}
\label{sec:ordering}
\begin{figure}
    \centering
    \includegraphics[scale=1]{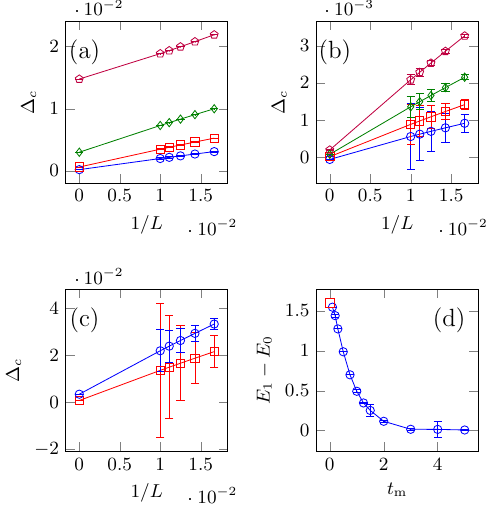}
    \caption{
    Charge gap $\Delta _{\rm c}(L)$ in regime 1
    of the (a) microscopic and (b) effective models with $L=100,t_{\rm m}=1.0$. 
    \textbf{(a)} $t_{\perp}=0.5$ (circle), $t_{\perp}=0.55$ (square), $t_{\perp}=0.6$ (diamond) , $t_{\perp}=0.65$ (pentagon).
    \textbf{(b)} $J_{\perp}=0.25$ (circle), $J_{\perp}=0.3025$ (square), $J_{\perp}=0.36$ (diamond) , $J_{\perp}=0.4225$ (pentagon). 
    \textbf{(c)} $\Delta _{\rm c}$ in regime 2 for the microscopic model (circle, $t_{\perp}=3.0$) and the effective model (square, $J_{\perp}=3.6$)  with $L=100,t_{\rm m}=10.0$.
    \textbf{(d)} The gap $E_{1}-E_{0}$ (difference between  first excited and ground state energy ) against $t_{\rm m}$  with $t_{\perp}=1.2$, $L=100$ in regime~1.
    Red square marks the gap from exact calculation of an isolated dimer, $t_{\rm m}=0$.
    }
    \label{fig 2}
\end{figure}
For $\hat{H}_{\rm KBP}$, we have performed numerical many-body calculations for two regimes: 
\begin{align}
(1) \quad U & = 4.0,\quad  1.0\leq t_{\rm m} \leq 5.0\\
(2) \quad U & = 10.0,\quad  4.0\leq t_{\rm m} \leq 12.0
\end{align}
In both regimes, a range of different $t_\perp$-values were run, and in some select cases the ground state properties of $\hat{H}_{\rm KBP}^{J_\perp}$ were computed for matching $J_\perp$-values.
Regimes~1 and 2 approximate the two regimes that we investigated in ref.~\cite{JEEBOT2025}, where pairing sites also had direct nearest-neighbor tunneling in-between them.
Regime~1 in that prior work, where $t_{\rm m}$ had been fixed to $1.0$, served as a 1D-version of the `conventional' proposals for reservoir-mediated stabilization of superconductivity~\cite{Erez2008,Gideon2012}.
Regime~2 in ref.~\cite{JEEBOT2025}, with $t_{\rm m}$ fixed to $10.0$, then was designed to simultaneously boost both pairing energy and superconducting stiffness - coupled with a systematic relative detuning of the nominal Fermi points of the pairing and the metallic layer - much more comprehensively  than previous versions of the proposal.
As the direct tunneling between pairing sites considered in~\cite{JEEBOT2025} complicates analytical understanding and precludes us from our aim of mapping the field of reservoir-enhanced superconductivity to that of Anderson-/Kondo-lattices, our pairing layer is simplified to consist only of~\cref{eq:hu}.

For both regimes we show examples of the pair-pair correlation functions $C_\lambda(i,j)$ and the modulus of the density-density correlations $|N_\lambda(i,j)|$ in~\cref{fig 1} for $\hat{H}_{\rm KBP}$ at $L=200$, both within the pairing layer (left-hand column), as well as within the metallic layer (right-hand column).
For the data shown in~\cref{fig 1}, as well as for any other calculation that we ran, we find the equality ${2C_\lambda(i,j)=|N_\lambda(i,j)|}$ to be fulfilled exactly to within numerical precision.
This degenerate behaviour of the pair-pair and the density-density correlations stands in contrast to other 1D and quasi-1D systems such as e.g. doped Hubbard-chains or Hubbard-ladders, whose algebraic decay at zero temperature is controlled by exponents proportional to $K^{-1}$ and $K$ respectively, where  $K$ denotes the TLL-parameter of the relevant charge mode.
The degeneracy of ${2C_\lambda(i,j)}$ and ${N_\lambda(i,j)}$ is the direct consequence of the $SU(2)$ spin-rotation symmetry of $\hat{H}_{\rm AL}$ at half-filling (naturally translating to $\hat{H}_{\rm KL}$), which implies that all spin-spin correlators ${\langle\hat{S}^a_{i,\lambda}\hat{S}^a_{j,\lambda}\rangle}$, ${a=x,y,z}$ are equal to each other.
Using the particle-hole mapping, the $SU(2)$-invariance of $\hat{H}_{\rm AL}$ and $\hat{H}_{\rm KL}$ translates to the observed degeneracy of pair-pair and density-density correlations for $\hat{H}_{\rm KBP}$ and $\hat{H}_{\rm KBP}^{J_\perp}$ (c.f.~\cref{app:degen}), with ${\langle\hat{S}^+_{i,\lambda}\hat{S}^-_{j,\lambda}\rangle}$ mapping to $C_\lambda(i,j)$ and $4{\langle\hat{S}^z_{i,\lambda}\hat{S}^z_{j,\lambda}\rangle}$ mapping to $|N_\lambda(i,j)|$.
\Cref{fig 1}a and c further illustrate how correlations may appear to be barely decaying at sufficiently low coupling for both regimes, even for the largest system sizes amenable to DMRG-type numerics, especially within the pairing layer.
At the same time, \cref{fig 1}e also illustrates that raising the coupling can rapidly lead to correlations exhibiting clear exponential decay across the system.

The scaling of the charge-gap $\Delta_c$ with $t_\perp$ or $J_\perp$, for $\hat{H}_{\rm KBP}$ and $\hat{H}^{J_\perp}_{\rm KBP}$ respectively, now acquires particular importance for understanding whether the results shown in~\cref{fig 1}c and e are explainable as a crossover process, wherein the length-scale associated with $\Delta_c$ shrinks exponentially as $\Delta_c$ grows exponentially with rising coupling in-between pairing and metallic layer.  
In order to investigate this hypothesis, we pursue two tracks, the first being an analytical renormalization group (RG) approach that allows us to tap into prior results for systems related under the RG flow~\cite{Aristov2010}, while the second one is the numerical track.

For the analytical approach, we first establish that the Hamiltonian $\hat{H}^{J_\perp}_{\rm KBP}$ at half-filling will generate a finite spin gap at any non-zero value of $J_\perp$, as detailed in \cref{app:rg}.
The resulting RG-flow leads to an effective model that maps to the asymmetric two-leg spin$1/2$ ladder when reversing the particle-hole mapping~\cref{eq:phtrafo}
For these systems, multi-method analysis has already shown that this effective system exhibits a finite gap at any coupling between the legs of the ladder~\cite{Aristov2010}.
When mapped back to $\hat{H}^{J_\perp}_{\rm KBP}$, this would exactly result in the charge gap scaling as ${\Delta_c\propto e^{-A/J_\perp}}$.

For the numerical track investigating $\Delta_c$, we compute $\Delta_c$ according to~\cref{eq:cgap} at ${L=60,70,80,90,100}$ for $\hat{H}_{\rm KBP}$ at half-filling, which has not been studied previously, then extrapolate $\Delta_c(L)$ to the thermodynamic limit $L\rightarrow\infty$.
We do so both for regime~1 as well as regime~2, and for each chosen value of $t_\perp$ we also compute $\Delta_c(L)$ for $\hat{H}_{\rm KBP}^{J_\perp}$ at matching $J_\perp$-values, i.e. for $J_\perp=4t_\perp^2/U$.
The results are shown in~\cref{fig 2}.
In~\cref{fig 2}a, we show $\Delta_c(L)$ and its extrapolated values for the 1D Kivelson bilayer proposal at half-filling for regime~1, for ${t_{\rm m}=1.0}$ and ${t_\perp=0.5,0.55,0.6,0.65}$.
These results are to be compared with the charge gaps obtained for the perturbative approximation $\hat{H}_{\rm KBP}^{J_\perp}$ at the same $t_{\rm m}$ and corresponding values of ${J_\perp=4t_\perp^2/U}$, ${J_\perp=0.25,0.3025,0.36,0.4225}$, as shown in~\cref{fig 2}b.
In~\cref{fig 2}c, analogous calculations for regime~2 are shown, namely $\Delta_c(L)$ and its extrapolated values at ${t_{\rm m}=10.0}$ and for ${t_\perp=3.0}$ and ${J_\perp=3.6}$ respectively.
Across all cases, we find that the charge-gap obtained for the approximative perturbative model underestimates the true value, and quite severely so in the case of regime~1.
Further, analysis of the scaling behaviour of the extrapolated values ${\Delta_c(L\rightarrow\infty)}$ shown in~\cref{fig 2}a with $t_\perp$ could be compatible with $e^{-B f(t_\perp^{-1})}$, with $B$ another constant depending on the microscopic physics, and $f(x)$ representing any low-order polynomial or power-law in $x$.
As the particle-hole mapping~\cref{eq:phtrafo} is exact, these results directly translate to the behaviour of the spin gap for the half-filled 1D Anderson- and Kondo-lattices respectively.
%
%
\begin{figure}
    \centering
    \includegraphics[scale=1]{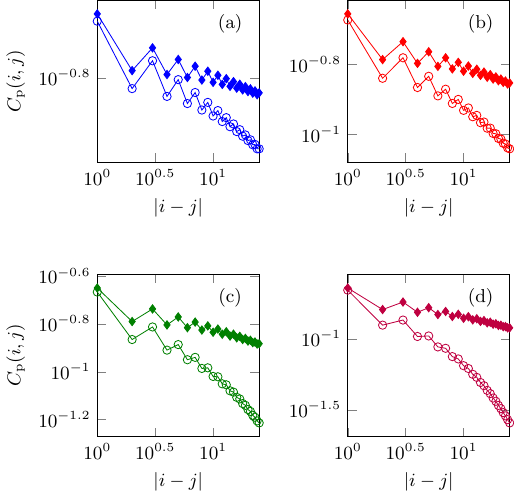}
    \caption{
    Comparison of pair-pair correlation function between microscopic model (open circles) and effective model (full diamonds) in regime~1 with $L=100,t_{\rm m}=1.0$. 
    \textbf{(a)} $t_{\perp}=0.5, J_{\perp}=0.25$, \textbf{(b)} $t_{\perp}=0.55,J_{\perp}=0.3025$, \textbf{(c)} $t_{\perp}=0.6,J_{\perp}=0.36$ and \textbf{(d)} $t_{\perp}=0.65,J_{\perp}=0.4225$. 
    } 
    \label{fig:pair_corr_regime1}
\end{figure}

The source of the ultimately fully gaped, and thus insulating nature of these systems in the thermodynamic regime is seen most straightforwardly in the limit in which $t_{\rm m}$ becomes small relative to $t_\perp$ and $U$, ${t_{\rm m}/t_\perp\rightarrow 0}$, ${t_{\rm m}/U\rightarrow 0}$.
In this limit, a pairing site and the metallic site to which it is coupled by $t_\perp$ form an effective dimer.
The system can thus be seen as a chain of coupled dimers and this coupling is weak when $t_{\rm m}$ is small compared to $t_\perp$ and $U$.
Moreover, at half filing, where $\mu_p=-U/2$ and $\mu_m=0$, the energetic cost of of having a doublon on the pairing site of a dimer and placing one on the metallic site of the dimer are degenerate.
Thus, for the isolated dimers symmetric and antisymmetric superpositions of doublon states, ${|\uparrow\downarrow,0\rangle}$, ${|0,\uparrow\downarrow\rangle}$ are prominent components of the low-lying eigenstates of the isolated doublons at half-filling.
For the half-filled groundstate, the amplitudes associated with single occupancies, ${|\uparrow,\downarrow\rangle}$, ${|\downarrow,\uparrow\rangle}$, moreover increase as ${t_\perp/U}$ increases, while ${|\uparrow\downarrow,0\rangle}$ and ${|0,\uparrow\downarrow\rangle}$ enter the ground state symmetrically.
For the first excited state, the amplitudes of configurations ${|\uparrow,\downarrow\rangle}$ and ${|\downarrow,\uparrow\rangle}$ are zero, while ${|\uparrow\downarrow,0\rangle}$ and ${|0,\uparrow\downarrow\rangle}$ enter the wave function anti-symmetrically (bonding and anti-bonding level for the effective ``boson'' formed by a pair).
The gap between these two states, which map to the singlet and triplet states under the particle-hole transformation~\cref{eq:phtrafo} and has thus been designated as the neutral singlet gap~\cite{Yu1993}, is of order $J_\perp$.
In~\cref{fig 2}d, we show an example of the gap between ground and first excited states for a system with $L=100$ and a range of finite coupling $t_{\rm m}$ between the dimers.
As ${t_{\rm m}\rightarrow 0}$, we recover the gap of the isolated dimers, while increasing $t_\perp$ leads to an exponentially decaying gap between ground and first excited state. 
But, as our analysis of ${\Delta_c(1/L)}$ in conjunction with the known results for the Kondo chain has shown, this gap almost certainly never disappears, even in the thermodynamic limit, always resulting in a ground state of that can be roughly visualized as one of ultimately localized doublons in the bonding band provided by the $t_\perp$-component of $\hat{H}_{\rm KBP}$.
In that, our results show that it corresponds, via the particle-hole mapping, with the picture of the Kondo-insulator as being formed from localized singlets forming on the bonds between metallic sites and localized orbitals.
%
%
\begin{figure*}
    \centering
    \includegraphics[scale=1]{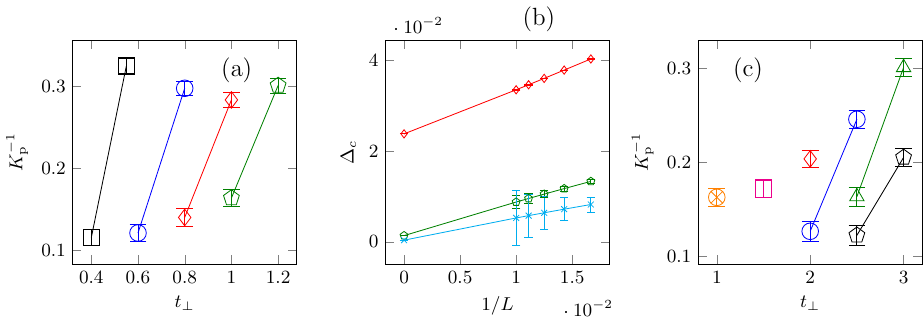}
    \caption{
    \textbf{(a)} $K^{-1}_{\rm p}$ vs. $t_{\perp}$ for $\hat{H}_{\rm KBP}$ in regime~1 at ${L=100}$. 
    $t_{\rm m}=1.0$ (square), $t_{\rm m}=2.0$ (circle), $t_{\rm m}=3.0$ (diamond) and $t_{\rm m}=4.0$ (pentagon).
    \textbf{(b)} $\Delta _{\rm c}(L)$ for $\hat{H}_{\rm KBP}$ at ${t_\perp=1.2}$.
    $t_{\rm m}=3.0$ (diamond), $t_{\rm m}=4.0$ (pentagon),  $t_{\rm m}=5.0$ (x).
    \textbf{(c)} $K^{-1}_{\rm p}$ vs. $t_{\perp}$ for $\hat{H}_{\rm KBP}$ in regime~2 at ${L=100}$. 
    $t_{\rm m}=2.0$ (otimes), $t_{\rm m}=4.0$ (square), $t_{\rm m}=6.0$ (diamond), $t_{\rm m}=8.0$ (circle), $t_{\rm m}=10.0$ (triangle) and $t_{\rm m}=12.0$ (pentagon).
    }
    \label{fig:Luttiger_liquid_par}
\end{figure*}

In summary, our findings show that the charge gap obtained for $\hat{H}_{\rm KBP}^{J_\perp}$ - and thus also the spin gap for $\hat{H}_{\rm KL}$ - are renormalized upwards by the higher-order terms that are usually dropped when deriving the effective Hamiltonian from $\hat{H}_{\rm KBP}$ - or from $\hat{H}_{\rm AL}$.
But with the available data and even with the achievable accuracy, we cannot presently distinguish between scaling as ${\Delta_c\propto e^{-B/t_\perp}}$, or as ${\Delta_c\propto e^{-B/t_\perp^2}}$, or even as the exponential of some low-order fractional power-law or polynomial of $t_\perp^{-1}$.
Even with this uncertainty, we find that the microscopic model only strengthens the physics that had been analytically derived for $\hat{H}_{\rm KL}$ and $\hat{H}_{\rm KBP}^{J_\perp}$, with a scaling of the gap that is exponential in a low-order polynomial or power-law of $t_\perp^{-1}$.
We thus conclude that the apparent onset of exponential decay of both ${C_\lambda(i,j)}$ and ${N_\lambda(i,j)}$ with rising $t_\perp$ is in fact a crossover, wherein the gap is exponentially small at small $t_\perp$.
Therefore, at small $t_\perp$, the exponentially decaying envelope of all correlation functions can easily become undetectably large even with the highest $L$-values amenable to current many-body numerics.
Our results also show that below the scale at which the exponentially decaying envelope is detectable, superconducting and density-density correlations are well-fitted by power-laws, which may moreover exhibit such low exponents as to seemingly approach long-range order.
We had previously found such near-long-range order for Kivelson's bilayer proposal in 1D, $\hat{H}_{\rm KBP}$, away from half-filling and with added nearest-neighbour tunneling within the pairing layer~\cite{JEEBOT2025}.
Different from the present work, raising $t_\perp$ did not give rise to crossover behaviour at any $L$ amenable to numerics in ref.~\cite{JEEBOT2025}, with pair-pair correlations staying algebraic throughout, even though their algebraic exponent and overall amplitude worsen markedly in the process.
We find the same to be true for $\hat{H}_{\rm KBP}$ away from half-filling, as we explore in separate work~\cite{JEEBOT2025_2}.
Taken together, these findings highlight the half-filled regime as the ultimate cause of the non-vanishing gap at any finite value of $t_\perp$.
Even so, at weak coupling between pairing and metallic layer, at half filling these systems can look nearly ordered on large length scales, both with respect to a superconducting as well as to a charge-density wave phase.
Mapped to the repulsively interacting models ${\hat{H}_{\rm AL}}$ and ${\hat{H}_{\rm KL}}$, this behavior corresponds to a near-AFM order on large scales, which is naturally degenerate in the three spatial directions.
While half-filling ultimately scuppers the prospect of ordering, these systems thus still exhibit a strong drive towards AFM order at intermediate size scales for weaker interlayer coupling.
In contrast, for the 2D Kondo lattice at sufficiently low $J_\perp$, AFM order is not just realized at intermediate scale, at but also in the thermodynamic limit~\cite{Assaad1999,Eder2016}.

In~\cref{fig:pair_corr_regime1}, we illustrate the crossover behaviour at it occurs in real space, as seen for $C_{\rm p}(i,j)$, for regime~1.
Specifically, for ${t_{\rm m}=1.0}$ and ${L=100}$, this data demonstrates how the correlation function changes markedly within a narrow window of $t_\perp$-values for $\hat{H}_{\rm KBP}$, ${t_\perp\in[0.5,0.55,0.6,0.65]}$, from all-algebraic to all-exponential.
This is contrasted with the behaviour of the same correlation function computed for the ground state of $\hat{H}_{\rm KBP}^{J_\perp}$ at matching values of $J_\perp$, which not only remains algebraically decaying at the scale of the system, but for which the algebraic exponent assigned from fitting~\cref{eq:cfit} also barely changes.
In that, it mirrors the physics of the charge gaps shown in~\cref{fig 2}a and b for the same systems, where $\Delta_c$ is far larger, and increases much more rapidly, for the full Kivelson bilayer proposal than for its approximative second-order perturbative version.
With $\Delta_c$ being exponentially small at low values of $t_\perp$, \cref{fig 1}a and~\cref{fig:pair_corr_regime1}a in particular illustrate that even the largest magnitudes of systems amenable to reliable many-body numerics may appear to look near-long-range-ordered, i.e. exhibit very slowly decaying power-law-like behaviour, for all practical purposes.
We find the same to hold in regime~2, c.f.~\cref{fig 2}c and~\cref{app:pair_pair}.
Depending on the exact magnitude of $\Delta_c(L)$, and the $L$-value for which $\hat{H}_{\rm KBP}$ is realized in a given material or quantum simulator, the same could as well be true in an experiment (c.f.~\cref{sec:disc_and_outl}).
In~\cref{fig:Luttiger_liquid_par}a and c we therefore show TLL-parameters $K_{\rm p}^{-1}$, obtained for a range of Hamiltonian parameters $t_\perp$ and $t_{\rm m}$ for regimes~1 and 2 at ${L=100}$, that we can assign to any system that exclusively exhibits power-law decay across the entire bulk of the lattice, i.e. within its center-half ${L/4\leq i \leq 3L/4}$, away from boundary effects.
As in our previous work~\cite{JEEBOT2025}, this data clearly shows that even these most simplified versions of the Kivelson bilayer proposal can achieve far better superconducting susceptibilities than the isolated pairing layer with short-range couplings, which are limited to $K_{\rm p}^{-1}=1/2$.
It also demonstrates how increasing $t_{\rm m}$ at constant $t_\perp$ yields significant enhancements to $K_{\rm p}^{-1}$ and thus superconducting susceptibilities.
Following the arguments laid out in~\cref{sec:rkky}, we interpret this effect as a reduction of the spin gap induced by the pairing layer inside the metallic layer, as kinetic energy is increased relative to induced pair binding.
In turn, the reduction in induced spin gap translates to an increased coherence length $\xi_{S,m}$ for single-particle propagation inside the metal, which we identify as the key driver for boosting superconducting susceptibility in~\cref{sec:rkky}.
Where lines terminate with growing $t_\perp$ in~\cref{fig:Luttiger_liquid_par}a and c, this is because the correlation functions stop behaving as pure power laws within the systems' bulk beyond this point.
Comparing~\cref{fig:Luttiger_liquid_par}a with~\cref{fig:Luttiger_liquid_par}b, illustrates this:
As $t_{\rm m}$ is increased from $3.0$ to $4.0$ at $t_\perp=1.2$, the charge gap ${\Delta_c(L\rightarrow\infty})$ decreases by a factor of about $17.0$.
This pushes the exponentially decaying envelope of $C_p(i,j)$ so far out that we can fit it with a power law,~\cref{eq:cfit} at ${L=100}$, which was not possible at $t_{\rm m}=3.0$, leading to the termination of the line for $t_{\rm m}=3.0$ beyond ${t_\perp=1.0}$ in~\cref{fig:Luttiger_liquid_par}a

Conversely, where lines terminate with decreasing $t_\perp$, this is due to persistent challenges in attaining converged ground states.
Even with high-performing implementations of two-site DMRG at large bond dimension (c.f.~\cref{app:sims}), low $t_\perp/t_{\rm m}$-ratios lead to persistent trapping in metastable excited states even at ${L=100}$, that the quasi-local DMRG updates are unable to overcome.
Extending these parameter-lines to lower $t_\perp/t_{\rm m}$-ratios thus would require either working at substantially larger values of $L$, where the reduced finite-size gaps enable conventional DMRG to escape the trap of metastable states even when $t_\perp$ is very small w.r.t.~$t_{\rm m}$, or deploying advanced techniques to escape the metastability-trap~\cite{Stoudenmire2012}. 
%
%

\begin{figure}[t!]
       \centering
     \hspace*{-0.5cm} 
      \includegraphics[scale=1]{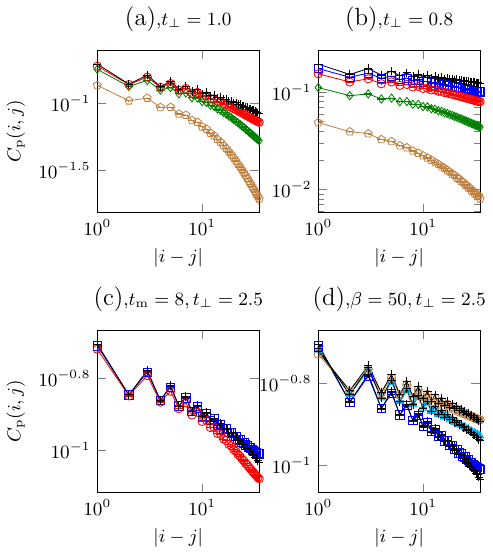}
       \caption{ 
       Pair correlation function of $\hat{H}_{\rm KBP}$ at finite temperature at ${L=100}$, and comparison with corresponding DMRG results ($+$).
       \textbf{(a $\&$ b)}: regime~1 at ${t_{\rm m}=3.0}$, ${\beta =90.0}$ (square), ${\beta =70.0}$ (circle), ${\beta =50.0}$ (diamond), ${\beta =30.0}$ (pentagon).
       %
       \textbf{(c $\&$ d)}: regime~2. 
       \textbf{(c)} ${t_{\rm m}=8.0}$, ${\beta =50.0}$ (square), ${\beta=30.0}$ (circle)  
        \textbf{(d)} ${t_\perp =2.5}$, ${\beta =50.0}$, ${t_{\rm m} =8.0}$ (square) ,  ${t_{\rm m} =10.0}$ (triangle),
       ${t_{\rm m} =12.0}$ (circle).
       }
       \label{fig:finite_temp}
   \end{figure}

Finally, we employ the AFQMC-based ALF-package~\cite{10.21468/SciPostPhysCodeb.1-v2.4,10.21468/SciPostPhysCodeb.1-r2.4} 
in order to study an example as to how the appearance of near-long-range order builds up with lowering temperature.
Shown in~\cref{fig:finite_temp} is $C_{\rm p}(i,j)$ at different values of inverse temperature ${\beta=1/T}$ in both regimes~1 and 2, for different combinations of $t_\perp$- and $t_{\rm m}$-values.
Each time, the DMRG results at $T=0$ are shown for reference.
At around $\beta=90.0$ at most, $C_{\rm p}(i,j)$ computed for the finite-temperature system largely matches the zero-temperature correlator in both magnitude and long-distance behaviour.
In regime~2, we find near-perfect agreement already from around $\beta=50.0$ onwards, while for regime~1 at least the long-range behaviour already largely coincides with that of the ground state from that point.
Depending on how exactly these systems were to be realized in an experiment (c.f.~\cref{sec:disc_and_outl}), the effective zero-temperature behaviour for finite clusters of significant size would thus be accessible at accessible non-zero values of $T$.
\section{The RKKY interaction from quasi-exact numerics in the dense regime}\label{sec:rkky}
\begin{figure}
\centering
    \includegraphics[scale=1]{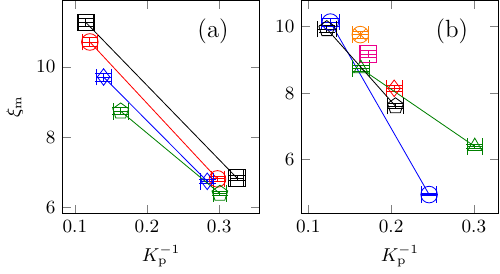}
    \caption{
    The single-particle propagation length scale $\xi _{\rm m}$ in the metallic layer vs. the TLL-parameter $K_{\rm p}^{-1}$.
    \textbf{(a)} Regime~1: $t_{\rm m}=1.0$ (square), $t_{\rm m}=2.0$ (circle), $t_{\rm m}=3.0$ (diamond) and $t_{\rm m}=4.0$ (pentagon).
    \textbf{(b)} Regime~2: $t_{\rm m}=2.0$ (otimes),$t_{\rm m}=4.0$ (square), $t_{\rm m}=6.0$ (diamond), $t_{\rm m}=8.0$ (circle), $t_{\rm m}=10.0$ (triangle) and $t_{\rm m}=12.0$ (pentagon).
    }
    \label{fig:rkky}
    \end{figure}

Spin-spin interactions mediated by the metal have been studied extensively for Kondo-lattices at different spatial dimensionality $D$~\cite{Rusin2017,Nejati2017}.
Known as Ruderman-Kittel-Kasuya-Yoshida (RKKY) interactions, their analytical treatment necessitates approximating the localized spins in the dilute limit, where only the two interacting spins are retained.
Thus, any back-action of the spin-lattice onto the metal is effectively neglected by construction.
These treatments yield spin-spin coupling that decays as a power-law in the distance of the two localized spins, with an algebraic exponent ${\alpha=2^{D-1}}$ for ${D=1,2,3}$, and thus these are especially long-ranged in the case of either one- or two-dimensional metals.

In transmuted form, these analytical results for the RKKY-interaction reappeared, independently and without mutual awareness, in work on the Kivelson bilayer proposals.
Specifically, refs.~\cite{Lobos2009,Gideon2012} showed that even a 2D metal with static disorder would mediate a pair-pair coupling inside a 1D wire with intrinsic pairing to which it weakly couples.
This coupling was predicted to decay quadratically with distance in imaginary time, and to the fourth power with spatial distance in the best case (i.e. on length scales below the metal's coherence length).
The effective long-range pair-pair coupling that the metal is thus predicted to mediate within the 1D wire in this approach was moreover shown to be sufficient to establish true superconducting long-range order, at least at zero temperature, apparently contravening the MWT.
However, as in the analytical study of the RKKY-interaction, it is difficult to incorporate the effects of the interacting layer onto the metallic layer.

In ref.~\cite{JEEBOT2025} that we have shown the impact of this back action onto the metal to be crucial.
Specifically, ref.~\cite{JEEBOT2025} demonstrated that the proximity-effect emanating from the pairing onto the metallic layer opens up a spin gap for the metal, which leads to a strong renormalization of the way single-particle excitations decay with distance in the metal, from algebraic in the free metal at ${t_\perp=0}$, to exponential at ${t_\perp>0}$.
With the metal-mediated pair-pair coupling becoming effectively short-ranged in this way, the MWT is ultimately restored, in that $C_{\rm p}(i,j)$ decays algebraically. 
At the same time, the mediated coupling can still effectively range across many sites.
As shown for the equivalent of regime~2 in that publication, the localization length $\xi_{S,m}$ extracted from fitting an exponentially decaying envelope function ${A_{S,m}e^{-|i-j|/\xi_{S,m}}}$ to the maxima of the oscillating single-particle density matrix ${S_m(i,j)}$ is inversely correlated with $K_{\rm p}^{-1}$.
That is, as the pair-pair coupling mediated by the metal increases in range, the superconducting susceptibility of the pairing layer $\chi(k,\omega,T)$, which scales as ${\max[k,\omega,T]^{K_{\rm p}^{-1}-2}}$~\cite{Thierrybook2003}, improves markedly, resulting in power-law decays so slow that they only become noticeable for very large systems - i.e. near-long-range-order.
In that, it is loosely confirming the results of ref.~\cite{Lobos2009}: in a system with intrinsic pairing which is dominated by pair-phase fluctuations (regime~2 in ref.~\cite{JEEBOT2025} is chosen to represent that regime), the longer-ranged the pair-pair coupling mediated by the metal, and thus the more the superconducting phase is stabilized.
Where refs.~\cite{Lobos2009} and~\cite{JEEBOT2025} diverge, is that our non-perturbative many-body numerics show that, at least in the case where both pairing and metallic layers are 1D and of a comparable density, the metallic layer can never mediate truly long-ranged, i.e. low-exponent algebraically decaying pair-pair coupling, and thus at most near-long-ranged order is attainable, but not true long-range order.

As we had previously obtained these results for ${\hat{H}_{\rm KBP}}$ away from half-filling and with added nearest-neighbour-tunneling between the pairing sites, here we demonstrate the same effect at half-filing and for ${\hat{H}_{\rm KBP}}$ alone.
As illustrated in~\cref{app:spdm}, ${S_m(i,j)}$ decays exponentially at any value of ${t_\perp>0}$ for these systems as well.
This general form of the single-particle density matrix in the metallic layer means that we can once more assign a length scale $\xi_{S,m}$ from fitting an exponentially decaying function to the maxima of ${S_m(i,j)}$.
This length scale quantifies the distance across which the single-electron excitations of the metallic layer can coherently mediate the pair-pair coupling within the pairing layer: as pairs enter the metal from the pairing layer, its constituents may either propagate independently through the metal and then re-enter the pairing layer after crossing a distance on the order of $\xi_{S,m}$ (c.f.~\cite{Lobos2009}).
There is of course an alternative channel by which the metal may attempt to stabilize superconductivity here, which is by propagation of bound electron pairs, which is now also possible due to the spin-gap induced into it by the pairing layer, i.e. due to the proximity effect - we study this separate effect in more depth in~\cite{JEEBOT2025_2}.

The effectiveness and power of the single-electron excitation channel in stabilizing the superconducting properties of the pairing layer, in line with Kivelson's basic idea, is borne out by plotting $\xi_{S,m}$ vs. $K_{\rm p}^{-1}$, as we do in~\cref{fig:rkky}.
As we had found for regime~2 of ref.~\cite{JEEBOT2025}, $\xi_{S,m}$ and $K_{\rm p}^{-1}$ are anti-correlated, but now both for regime~1 and regime~2.
This highlights the fact that dropping the direct tunneling between pairing sites pushes regime~1 of the present work to also be dominated by phase- instead of amplitude-fluctuations, as regime~1 in ref.~\cite{JEEBOT2025} had been.
Thus, increasing the range of metal-mediated coupling always enhances the superconducting susceptibility.

With this result established, we can directly transfer it to $\hat{H}_{\rm KBP}^{J_\perp}$, $\hat{H}_{\rm AL}$ and $\hat{H}_{\rm KL}$, all at half-filling, via Schrieffer-Wolff transformation, by using~\cref{eq:phtrafo}), or by combining the two.
Our findings thus immediately translate to the RKKY-interaction, which we hereby show to decay exponentially with distance for the 1D Kondo-necklace when the density of orbitals is comparable to that of the metallic sites.
This is due to the back-action of the localized orbitals onto the metal, which induces a small effective charge gap inside the metallic layer.
This induced gap is the cause of the RKKY-interactions becoming short-ranged, and the system's inability to either reach true long-range order, or even just to escape the fully insulating state in the thermodynamic limit that the half-filling condition can thus impose.

\section{Discussion \& Outlook}\label{sec:disc_and_outl}
This work shows that at half-filling neither the simplest 1D version of Kivelson's bilayer proposal nor the 1D Anderson lattice can escape having fully gapped insulating ground states in the thermodynamic regime, as had originally just been derived for the 1D Kondo necklace, where coupling between the layer of localized orbitals and the metallic layer had been approximated using second-order perturbation theory.
In fact, by directly treating the microscopic tunneling of electrons between the two layers, our data indirectly demonstrates that the higher orders of the perturbative expansion, which are usually not considered, only strengthen the smaller of the two gaps of these systems, that is the charge gap (on the negative-$U$ side, for $\hat{H}_{\rm KBP}$), respectively the spin gap (on the positive-$U$ side, for $\hat{H}_{\rm AL}$).
At the same time, our data also shows that as this gap is exponentially suppressed at small $t_\perp$, these systems not only can show appreciable superconducting and density-density correlations (for the Kivelson bilayer proposal) respectively spin-spin correlations (for the Anderson-lattice) over considerable distance, but ones that show such slow algebraic decay as to seemingly realizing near-ordered states.
We trace the inability of these half-filled 1D systems to escape the MWT and attain true long-range order, or even just to overcome the smaller of their two gaps, to the fundamental renormalization of the coupling mediated by the metallic layer, from algebraically to exponentially decaying.
In the terminology of the Kondo-necklace, we thus establish that the RKKY-interaction in fact is not long-ranged at all in these systems, but effectively short-ranged, due to the back-action of the pairing layer onto the metallic layer.

The present work has several implications for future research.
For one, it provides tools for merging the fields of superconductivity enhanced by metallic reservoir and Kondo physics, well beyond the present material.
As shown in~\cref{app:degen} the general Hamiltonian for the Kivelson bilayer proposal in the grand canonical ensemble, where each layer has its own chemical potential, maps to the Hamiltonian of the Anderson lattice, where the chemical potential terms for each layer now encode layer-specific magnetic fields; this mapping is independent of dimensionality and only presupposes a bipartite lattice.
In this way, studies of the Kivelson bilayer proposal for any given bipartite lattice and filling fractions in pairing and metallic layers simultaneously illuminate the physics of the periodic Anderson model on the same lattice at half filling with particular spin-imbalances in the orbital and metallic layers.
This exact mapping can thus be exploited in many different ways.
We are preparing work on the same 1D Kivelson bilayer proposal as studied here, but away from half-filling, the results of which will translate directly to the physics of the 1D Anderson lattice exposed to various magnetic fields~\cite{JEEBOT2025_2}.
Other applications of the mapping introduced here would be to e.g. re-interpret the quantum phase transition predicted for the 2D Kondo lattice at half-filling between AFM and insulator~\cite{Assaad1999,Eder2016}, as a transition between an ordered state with coexisting superconductivity and charge-density wave order and a fully gapped insulator for the 2D version of $\hat{H}_{\rm KBP}^{J_\perp}$.
In the opposite direction, 2D versions of Kivelson's bilayer proposal away from half-filling have been studied with a special focus on the Berezinskii-Kosterlitz-Thouless (BKT) transition temperature from a disordered state towards one with algebraically decaying superconducting quasi-order~\cite{Erez2008,Gideon2012}.
These predictions can thus be immediately translated to yield quantitative calculations for BKT transition temperatures for in-plane AFM quasi-order realized by 2D Anderson-lattices exposed to magnetic fields.

Experimental verification of the predictions made here could take multiple forms.
For one, with current quantum gas microscope set-ups, Hamiltonians such as~\cref{model_equ} or~\cref{eq:hal} could be engineered in a well-controlled manner, including the half-filling condition, e.g. by utilizing four instead of the more widely used two hyperfine states.
Two of these hyperfine states would need to be chosen to each have minimal or no scattering with the three other hyperfine states.
Then the tunneling between `pairing' and `metallic' layers could be achieved via coherently driven resonant Rabi oscillations between the two interacting hyperfine states (the `pairing layer') and the two non-interacting ones (the `metallic layer').
Any realization using ultra cold lattice gases would be predicated on the ability of the experiment to reach a sufficiently low entropy per particle, either via standard cooling, or by using adiabatic state preparation~\cite{Rabl2003,Kantian2010} adapted to these Hamiltonians, possibly combined with techniques to shuttle entropy into a sacrificial subsystem~\cite{Kantian2018,Yang2020,Xu2025}.
At current experimental capabilities, achievable $L$-values might range into the lower tens of sites.
Experimental verification in solid state state systems, most likely centred on the repulsive-$U$ side, might offer larger system sizes, as well as better established ways for reaching the low-temperature regime of these Hamiltonians.
This would come at the cost of increased effort to ascertain what microscopic Hamiltonian exactly has been realized.
Engineered chains of ad-atoms atop a metallic surface, with a thin insulating layer in between, close variants to systems that have already been realized~\cite{Neel2011}, would thus be another way to test our theoretical predictions.
Another way to probe the physical effects we studied here might be measurements of spin-spin correlations in heavy-fermion compounds realizing quasi-1D Kondo-lattices, such as CeCo$_2$Ga$_8$~\cite{Wang2017}.
Both spin-excitation spectra derived from high-resolution neutron-scattering~\cite{Shull1995,Qimiao2010,STOCKERT2023,Steglich2016}, or locally resolved magnetic susceptibilities recorded via NV-center magnetometry~\cite{Bonato2016,Arshad2024} would be able to observe the near-long-range order predicted to occur on intermediate size scales in this work.

Especially exciting is the prospect that experiments on the wide array of existing heavy-fermion compounds~\cite{Stewart1984}, when subjected to magnetic fields~\cite{Aoki2013}, could offer a way of indirectly testing the performance of Kivelson's bilayer proposal, in 1D, 2D and even 3D.
Thus, using magnetic fields in the range of a few Tesla, detecting and quantifying AFM orders in any such compounds with spin-exchange couplings on the order of $10^0$ to $10^1$ {K}, such as CeRhIn$_5$C~\cite{Das2014}, CeCo$_2$Ga$_8$~\cite{Wang2017}, CeSiI~\cite{Fumega2024} (and a host of other materials, as can be estimated from the N\'eel temperatures in Tab. 2 in ref.~\cite{Aoki2013}) would allow to indirectly study superconductivity stabilized by reservoir at a range of different filling factors.
Possibly, the different subtypes of AFM order found beyond some critical magnetic field in some compounds~\cite{Pourret2017} can be mapped, using~\cref{eq:phtrafo}, to a situation where coexistence between superconducting and CDW order ends, and only the former prevails.

Even without external magnetic fields to tune away from half-filling, the many AFM-like states occurring in 2D and 3D Kondo-lattices realized in heavy fermion materials~\cite{Aoki2013} could already be mapped to versions of the Kivelson bilayer proposal.
This offers the prospect of indirectly studying the competition between superconducting and CDW order based on already existing experimental data.
Such magnetically ordered states could map straight to superconductivity using~\cref{eq:phtrafo}.
\section*{Acknowledgements}\label{sec:acknow}
This work was supported by an ERC Starting Grant from the European Union’s Horizon 2020 research and innovation programme under grant agreement No. 758935; and the UK’s Engineering and Physical Sciences Research Council [EPSRC; grant numbers EP/W022982/1 and UKRI2088]. 
This work is also supported by the Swiss National Science Foundation under grant number 200020-219400.
The computations were enabled by resources provided through multiple EPSRC ``Access to HPC'' calls (Spring 2023, Autumn 2023, Spring 2024 and Autumn 2024) on the ARCHER2, Peta4-Skylake and Cirrus compute clusters, as well as by computer time awarded by the National Academic Infrastructure for Supercomputing in Sweden (NAISS).
This work was supported by a grant from the Swiss National Supercomputing Centre (CSCS) under project ID s1307 on Alps.
The authors also acknowledge the use of the HWU high-performance computing facility (DMOG) and associated support services in the completion of this work.

\appendix

\section{Pair and density correlations degeneracy}
\label[appendix]{app:degen}
\subsection{From attractive to repulsive interaction}
We start off with the Hamiltonian of the system
\begin{widetext}
\begin{equation}
    \begin{aligned}
        \hat{H}_{\rm KBP} &= -t_{\rm m} \sum_{i,\sigma}(\hat{c}^\dagger_{i,m,\sigma}\hat{c}^{}_{i+1,m,\sigma}+\text{h.c.})-U\sum_i \hat{n}_{i,p,\uparrow}\hat{n}_{i,p,\downarrow}-t_\perp\sum_{i,\sigma}(\hat{c}^{\dagger}_{i,p,\sigma}\hat{c}^{}_{i,m,\sigma} + \text{h.c.} ) \\  
        &\qquad \qquad - \mu_{p}\sum_{i}(\hat{n}_{i,p,\uparrow}+\hat{n}_{i,p,\downarrow}-1) - \mu_{m}\sum_{i}(\hat{n}_{i,m,\uparrow}+\hat{n}_{i,m,\downarrow}-1)
    \end{aligned}
    \label{eq:system_Hamiltonian}
\end{equation}
\end{widetext}
which corresponds to a one bath of free spin-$1/2$ fermions (operators $\hat{c}^\dagger_{i,m,\sigma}$ and chemical potential $\mu_m$) connected via a single-particle term $t_\perp$ to the pairing layer comprised of negative-$U$ centers with onsite attractive interaction (operators $\hat{c}^\dagger_{i,p,\sigma}$ and chemical potential $\mu_p$). 
The pairing layer is characterized by the absence of intrachain hopping terms. The indexes $i$ and $\sigma$ are those for sites in each layer and spin, respectively. 

We rewrite the interacting term as
\begin{widetext}
    
\begin{equation}
    \sum_j \hat{n}_{i,p,\uparrow}\hat{n}_{i,\downarrow} = \sum_i \left ( n_{i,p,\uparrow} -\frac{1}{2}\right ) \left ( \hat{n}_{i,p,\downarrow} -\frac{1}{2}\right ) + \frac{1}{2}\left ( \hat{n}_{i,p,\uparrow}+\hat{n}_{i,p,\downarrow} -\frac{1}{2}\right )
\end{equation}

\end{widetext}

to perform the following transformation. 
If we write the Hamiltonian in this form, it is straightforward to see that we can map the problem to repulsive local interaction by performing a double particle-hole transformation,~\cref{eq:phtrafo}, 
which results in taking $\hat{n}_{i,\lambda,\uparrow}\rightarrow \hat{n}_{i,\lambda,\uparrow}$ and $\hat{n}_{i,\lambda,\downarrow}\rightarrow 1-\hat{n}_{i,\lambda,\downarrow}$ in the metallic and pairing layers~\cite{Ho2009,Scalettar1989,Shun-Qing1996}. 
The interaction term changes its sign, $-U\rightarrow +U$ while the hopping terms remain unchanged
\begin{widetext}
    
\begin{equation}
    \begin{aligned}
        \hat{H}_{\rm AL} &= -t_{\rm m} \sum_{i,\sigma}(\hat{d}^\dagger_{i,m,\sigma}\hat{d}^{}_{i+1,m,\sigma}+\text{h.c.})+U\sum_i \left ( \hat{n}_{i,p,\uparrow} -\frac{1}{2}\right ) \left ( n_{i,p,\downarrow} -\frac{1}{2}\right )-t_\perp\sum_{i,\sigma}(\hat{d}^{\dagger}_{i,p,\sigma} \hat{d}^{}_{i,m,\sigma} + \text{h.c.})  \\
        & \qquad \qquad  - \frac{U}{2} \sum_i \left ( \hat{n}_{i,p,\uparrow}-\hat{n}_{i,p,\downarrow} +\frac{1}{2}\right ) -  \mu_{p} \sum_{i}(\hat{n}_{i,p,\uparrow} -\hat{n}_{i,p,\downarrow}) - \mu_{m} \sum_{i}(\hat{n}_{i,m,\uparrow} - \hat{n}_{i,m,\downarrow})
    \end{aligned}
\end{equation}

\end{widetext}
The chemical potential terms become now external magnetic fields which are non-zero only way from half-filling. 
Being at half-filling we have no external magnetic fields and we set
\begin{equation}
    \mu_{p}=-\frac{U}{2}, \quad \mu_{m}=0
\end{equation}
The repulsive Hamiltonian reads
\begin{widetext}
\begin{equation}
    \begin{aligned}
        \hat{H}_{\rm AL} &= -t_{\rm m} \sum_{i,\sigma}(\hat{d}^\dagger_{i,m,\sigma}\tilde{d}^{}_{i+1,m,\sigma}+\text{h.c.})+U\sum_i \left ( \hat{n}_{i,p,\uparrow} -\frac{1}{2}\right ) \left ( \hat{n}_{i,p,\downarrow} -\frac{1}{2}\right )-t_\perp\sum_{i,\sigma}(\hat{d}^{\dagger}_{i,p,\sigma} \hat{d}^{}_{i,m,\sigma} + \text{h.c.}) - \frac{UM_p}{4}
    \end{aligned}
\end{equation}
\end{widetext}
with $M_p$ the number of sites in the pairing layer.
\subsection{Mapping to spins: SU(2) symmetry}
We now consider the operators corresponding to s-wave superconductivity (SC) and the charge density wave (CDW) order. 
We perform the particle-hole transformation,~\cref{eq:phtrafo}, and map the fermionic operators onto spins as follows:
\begin{widetext}
\begin{equation}
    \begin{aligned}
        \hat{\Delta}^\dagger_{i,\lambda} \equiv \hat{c}^\dagger_{i,\lambda,\uparrow} \hat{c}^\dagger_{i,\lambda,\downarrow} &\overset{\text{p-h}}{\leftrightarrow}  (-1)^{i+\delta_{m,\lambda}}\hat{d}^\dagger_{i,\lambda,\uparrow} \hat{d}^{}_{i,\lambda,\downarrow} = \hat{S}^+_{i,\lambda} = \frac{1}{2}(\hat{S}^x_{i,\lambda}+i \hat{S}^y_{i,\lambda})\\
         \hat{c}^\dagger_{i,\lambda,\uparrow}\hat{c}^{}_{i,\lambda,\uparrow}+\hat{c}^\dagger_{i,\lambda,\downarrow}\hat{c}^{}_{i,\lambda,\downarrow}-1=\hat{n}_{i,\lambda,\uparrow}+\hat{n}_{i,\lambda,\downarrow}-1 &\overset{\text{p-h}}{\leftrightarrow} \hat{n}_{i,\lambda,\uparrow}-\hat{n}_{i,\lambda,\downarrow} = \hat{d}^\dagger_{i,\lambda,\uparrow}\hat{d}^{}_{i,\lambda,\uparrow}-\hat{d}^\dagger_{i,\lambda,\downarrow}\hat{d}^{}_{i,\lambda,\downarrow} = 2\hat{S}^z_{i,\lambda}\\
    \end{aligned}
\end{equation}
\end{widetext}

which means that the pair-pair and density-density correlations map directly to the spin-spin correlations
\begin{equation}
    \begin{aligned}
        \langle \hat{c}^\dagger_{i,\lambda,\uparrow} \hat{c}^\dagger_{i,\lambda,\downarrow} \hat{c}^{}_{j,\lambda,\downarrow} \hat{c}^{}_{j,\lambda,\uparrow}  \rangle &= (-1)^{i-j}\langle \hat{S}^+_{i,\lambda} \hat{S}^{-}_{j,\lambda} \rangle \\
        \langle ( \hat{n}_{i,\lambda} -1 ) ( \hat{n}_{j,\lambda} -1 ) \rangle &= 4\langle \hat{S}^z_{i,\lambda} \hat{S}^z_{j,\lambda} \rangle
    \end{aligned}
\end{equation}
The double particle-hole symmetry at half-filling gives a transformed model with no spin imbalance: this implies a global $SU(2)$ symmetry which corresponds to having
\begin{equation}
    \langle S^x_i S^x_{j} \rangle = \langle S^y_i S^y_{j} \rangle = \langle S^z_i S^z_{j} \rangle.
\end{equation}
For the initial attractive case, this corresponds to
\begin{equation}
    \langle \hat{\Delta}^\dagger_i \hat{\Delta}^{}_{j} \rangle = \frac{1}{2} \langle \left( \hat{n}_i -1\right)\left( \hat{n}_j -1\right) \rangle 
\end{equation}
To conclude, for the attractive Hubbard model at half-filling, CDW and s-wave SC correlations are exactly degenerate, up to a prefactor of $2$. 
Therefore, we expect the same scaling exponent at half-filling filling. 
Away from that filling-fraction, we generate effective magnetic fields that break the $SU(2)$-symmetry, and different correlations will generally show very different behaviour from one another.


\section{Single-electron correlations}
\begin{figure*}
    \centering
    \includegraphics[scale=1]{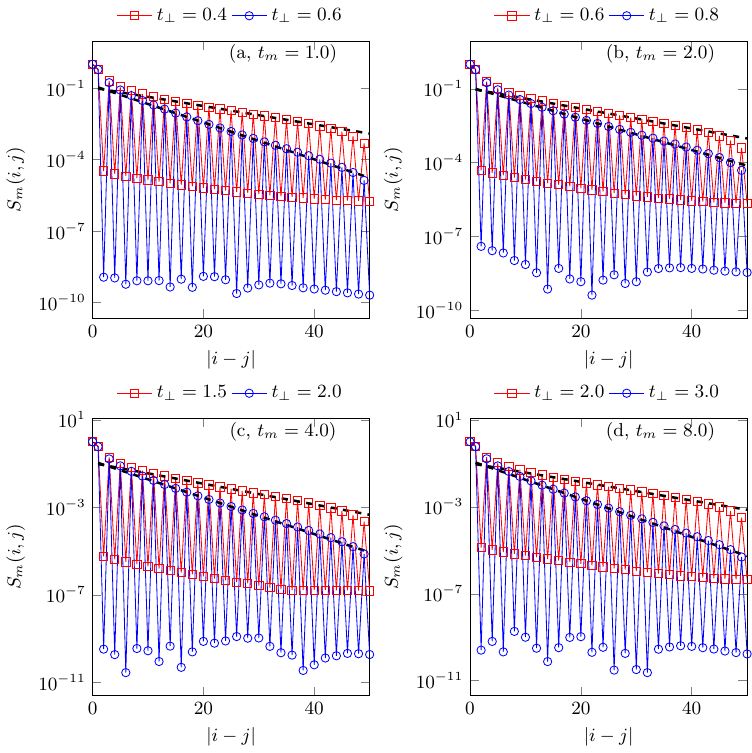}
    \caption{ 
    Single-particle correlation function $S_m(i,j)$. 
    {\bf{(a $\&$ b)}} Regime 1.
    {\bf{(c $\&$ d)}} Regime 2.
    }
    \label{fig:placeholder}
    \end{figure*}

\label[appendix]{app:spdm}
%
As discussed in~\cref{sec:rkky}, the back-action of the pairing onto the metallic layer fundamentally changes the physics of the latter.
Specifically, the single particle correlations, $\langle \hat{c}^\dagger_{i,m,\sigma}(\tau)\hat{c}^{}_{j,m,\sigma}(\tau')\rangle$, which would show decay with ${|i-j|^{-1}}$ and ${|\tau-\tau'|^{-1}}$ at ${t_\perp=0}$, now decay exponentially in these distances for ${t_\perp>0}$.
The reason for this is the spin-gap induced inside the metallic layer via the proximity-effect caused by the pairing layer.
In~\cref{fig:placeholder}, we show the resulting exponential decay of ${S_m(i,j)=\langle \hat{c}^\dagger_{i,m,\sigma}\hat{c}^{}_{j,m,\sigma}\rangle}$, which we consistently find at any ${t_\perp>0}$.
We also show the fits of the maxima of $S_m(i,j)$ to an exponentially decaying function, from which we extract the length-scale $\xi_{S,m}$, on which our analysis in~\cref{sec:rkky} is based.
\section{Pair correlations in regime 2}
\begin{figure}
    \centering
    \includegraphics[scale=1]{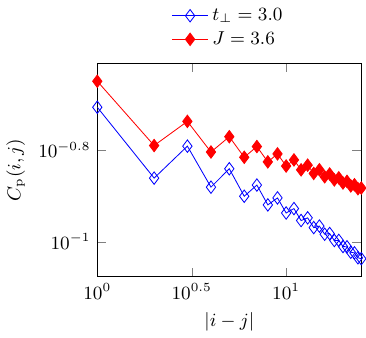}
    \caption{Pair-pair correlation function in regime 2 with ${L=100}$ for the microscopic model (open diamonds) and effective model (shaded diamonds).}
    \label{fig:placeholder_2}
\end{figure}

\label[appendix]{app:pair_pair}
%
In~\cref{sec:ordering}, we discuss the differences between the fully microscopic model described by Hamiltonian $\hat{H}_{\rm KBP}$, \cref{model_equ}, and the effective model based on the Schrieffer-Wolff-transformation combined with second-order perturbation theory, described by Hamiltonian $\hat{H}_{\rm KBP}^{J_\perp}$, \cref{eq:hkbpj}.
In~\cref{fig:pair_corr_regime1}, we illustrate the difference between these two models in the behavior of $C_p(i,j)$ for regime~1: 
as $\hat{H}_{\rm KBP}^{J_\perp}$ significantly underestimates the size of the charge gap $\Delta_c$, the pair-pair-correlators computed for these two different Hamiltonians increasingly diverge from each other as the coupling between pairing and metallic layers grows.
Here, we show the same to be the case for regime~2.
Specifically, \cref{fig:placeholder_2} contains $C_p(i,j)$ computed for the ground states of both $\hat{H}_{\rm KBP}$ and $\hat{H}_{\rm KBP}^{J_\perp}$ for ${L=100}$, at $t_\perp=3.0$ and $J_\perp=3.6$ respectively.
With $\Delta_c$ in the limit $L\rightarrow\infty$ extrapolating to a value roughly $4.5$ times larger in the former case compared to the latter, the apparent algebraic decay of the pair-pair-correlations in~\cref{fig:placeholder_2} is already markedly different between the microscopic and the effective models.
\section{Simulation parameters}
\label[appendix]{app:sims}
%
We have used open boundary conditions with two-site DMRG on~\cref{model_equ} and~\cref{eq:hkbpj} to measure observables at $T=0$ with $\mu _{\rm p} = 0$ and $\mu _{\rm m} = U/2$. 
We work at a bond dimension of around $8000$, and achieve truncated weights of about $\mathcal{O}(10^{-14})-$ $\mathcal{O}(10^{-16})$ as a result ~\cite{Schollwock2011}.

Finite temperature calculations with periodic boundary conditions are performed for~\cref{model_equ} using the Auxiliary field quantum Monte-Carlo (AFQMC) method as implemented in the ALF-package~\cite{10.21468/SciPostPhysCodeb.1-v2.4,10.21468/SciPostPhysCodeb.1-r2.4}. 
We use the standard $M_z$-decoupling which is known to yield stable results with low autocorrelation times for systems that only contain attractive interactions, such as the ones that we treat here.

\section{Spin gap from renormalization group analysis}
\label[appendix]{app:rg}
The goal of this section is to use the renormalization group (RG) approach to show that a pair hopping term favours the creation of a spin gap, even for a system that initially has no spin gap.

In the bosonization language \cite{Thierrybook2003}, we express phase and density excitations in terms of the fields $\theta(x)$ and $\phi(x)$ for both the spin (subscript $\sigma$) and charge (subscript $\rho$) sectors. For sufficiently large $U$, the p-layer is either doubly occupied or empty, we map it onto spin excitations with pseudospin $1/2$ (doubly occupied and empty states corresponding to $S^z=\pm 1/2$). The bosonization treatment consists of retaining only the low-energy physics and using the dictionary
\begin{equation}
    c^\dagger_{r,s} = \frac{1}{\sqrt{2\pi \alpha}}e^{irk_Fx}e^{-\frac{i}{\sqrt{2}}[r\phi_\rho-\theta_\rho+s(r\phi_\sigma-\theta_\sigma)]}
\end{equation}
where $r = \{R,L\}$ stands for right and left movers, $s=\{\uparrow, \downarrow\}$ and $\alpha$ is the short-distance cutoff. The Hamiltonian we bosonize is of the form
\begin{equation}
    \begin{aligned}
        \mathcal{H} &= \mathcal{H}_\text{metal} - 2J \sum_j\Big [S^\dagger_jc^\dagger_{j,\uparrow}c^\dagger_{j,\downarrow}+\text{H.c.} \Big] \\
        &\qquad + 4J \sum_j S^z_j(n_j-1)
    \end{aligned}
\end{equation}
and in terms of the bosonic fields it reads
\begin{equation}
    \begin{aligned}
        \mathcal{H} &= \mathcal{H}^{\rho,\sigma}_{0} 
        - J^{\pm}_f \sum_j\Big [S^+_j  e^{i\sqrt{2}\theta_\rho}\cos(\sqrt{2}\phi_\sigma)  + \text{H.c.} \Big] \\
        &\quad + J^{\pm}_b \sum_j\Big [S^+_j (-1)^je^{i\sqrt{2}\theta_\rho}\cos(\sqrt{2}\phi_\rho)  + \text{H.c.} \Big] \\
        &\quad \quad + J^{z}_b \sum_j S^z_j(-1)^j\cos(\sqrt{2}\phi_\rho) \cos(\sqrt{2}\phi_\sigma)  \\
        &\quad \quad \quad - J^{z}_f \sum_j S^z_j \nabla \phi_\rho \\
        &\quad \quad \quad \quad + \frac{y_\sigma}{2\pi\alpha} \sum_j \cos(\sqrt{8}\phi_\sigma)
    \end{aligned}
    \label{eq:bosonized_spin_fermions}
\end{equation}
where the first term describes the metallic layer (quadratic in the fields), the second term $J^\pm_f=\frac{2J}{\pi\alpha}$ refers to $c^\dagger_{r,s}c^\dagger_{-r,-s}$ while $J^\pm_b=\frac{2J}{\pi\alpha} $ refers to $c^\dagger_{r,s}c^\dagger_{r,s}$ and similarly for $J^z_f=\frac{4\sqrt{2}J}{\pi}$ with $c^\dagger_{r,s}c^{}_{r,s}$ and $J^z_b=\frac{8J}{\pi\alpha}$ with $c^\dagger_{r,s}c^{}_{-r,s}$. The last term $y_\sigma = g_\sigma/\pi\alpha$ 
is usually present when there are interactions. In our case this term is initially  zero ($g_\sigma =0$) but can be generated by renormalization. 
by $J^\pm_f$ and $J^\pm_b$ terms. As we have $K_\sigma=K_\rho=1$, all terms in the Hamiltonian have scaling dimension 2 and a priori are important on equal footing. Our goal is to show that $J^\pm_{f}$ and $J^z_b$ drive $g_\sigma$ towards strong coupling. As a consequence, the spin sector becomes gapped and we simplify the Hamiltonian by replacing $\cos(\sqrt{8}\phi_\sigma)$ with a constant $\mathcal{C}_\sigma \propto \langle \cos(\sqrt{8}\phi_\sigma )\rangle_{\mathcal{H}_\sigma}$. The contractions of the fields which give the terms contributing to $g_\sigma$ follow the idea of \cite{Giamarchi_Schulz_localization_PRB1988}.

The perturbative RG procedure in this case involves spin and fermionic operators. In the second order expansion in $J^\pm_f$, we approximate that $\langle S^+(x)S^-(x')\rangle \sim \frac{1}{\alpha}\delta(x-x')$ as the states over which we perform the average satisfy $\langle S^z\rangle=0$, meaning that there is no net magnetization in the p-layer. Our result strongly relies on such an approximation. The RG equation for the pair hopping and spin flip reads $\frac{dJ^{\pm}_f}{dl}=\frac{1}{2}(3-\frac{1}{K_\rho}-K_\sigma)J^\pm_f$. By looking at the scaling dimension of the operators, it follows also that $\frac{dJ^{z}_b}{dl}=\frac{1}{2}(3-K_\rho-K_\sigma)J^z_b$ and similarly for the other operators in \eqref{eq:bosonized_spin_fermions}. 

We start off by considering the contribution of $J^\pm_f$ to $g_\sigma$. For this purpose, we are interested in terms for which $|\textbf{r}-\textbf{r'}|< \alpha$, with $\textbf{r}=(x,u\tau)$ the space-imaginary time vector, then the corresponding averages read
\begin{widetext}
    \begin{equation}
    \begin{aligned}
        &\frac{(J^\pm_f)^2}{2u^2} \frac{y_\sigma}{\pi \alpha u }\int d^2\textbf{r}'\, d^2\textbf{r}'' \, d^2\textbf{r}''' \, \frac{\delta(x'-x'')}{\alpha}\langle \big (e^{i \sqrt{2}\theta_\rho}\cos (\sqrt{2}\phi_\sigma) + \text{H.c.} \big)(\textbf{r}') \big (e^{-i \sqrt{2}\theta_\rho}\cos (\sqrt{2}\phi_\sigma) + \text{H.c.} \big)(\textbf{r}'') \\
         &\quad \quad \quad \quad \quad \quad \quad \quad \quad\quad \quad \quad \times \cos (\sqrt{8}\phi_\sigma(\textbf{r}''')) \rangle_0 \\
         &=\frac{(J^\pm_f)^2}{2u^3} \frac{y_\sigma}{\pi \alpha^2 }\int d^2\textbf{r}'\, d\tau'' \, d^2\textbf{r}''' \, \langle \big (e^{i \sqrt{2}\theta_\rho}\cos (\sqrt{2}\phi_\sigma) + \text{H.c.} \big)(\textbf{r}')  \big (e^{-i \sqrt{2}\theta_\rho}\cos (\sqrt{2}\phi_\sigma) + \text{H.c.} \big)(x', \tau'') \\
         &\quad \quad \quad \quad \quad \quad \quad \quad \quad\quad \quad \quad \times \cos (\sqrt{8}\phi_\sigma(\textbf{r}''')) \rangle_0 \\
         &=\frac{(J^\pm_f)^2 y_\sigma}{u^3} \frac{d\alpha}{\pi \alpha^2}\int d^2\textbf{r}' \, d^2\textbf{r}''' \langle \cos (\sqrt{8}\phi_\sigma (\textbf{r}')) \cos (\sqrt{8}\phi_\sigma(\textbf{r}''')) \rangle_0 + \dots (x' \neq x'') \dots 
    \end{aligned}
\end{equation}
\end{widetext}
with the averages performed over $\mathcal{H}_0^{\rho,\sigma}$. Here, we replaced the spin operators by the a delta function and isolated the local contributions $\tau'' = \tau' + d\alpha$. By neglecting the last set of terms for $x'\neq x''$, we apply the standard RG procedure by changing cutoff to find the recursive equation which results in \begin{equation}
    \frac{d\tilde{y}_\sigma}{dl}=(2-2K_\sigma)\tilde{y}_\sigma - \frac{1}{2}(\tilde{J}^\pm_f)^2
\end{equation}
where the tilde refers to the corresponding dimensionless coupling. In an equivalent way, we also find that $(\tilde{J}^z_b)^2$ favours the creation of a finite $y_\sigma$ under the RG procedure. Indeed, even if we start with $K_\sigma=1$ (free fermions), we have that $d\tilde{y}_\sigma/dl\neq 0$ justify the simplification of considering the spin degrees of freedom to be frozen. For the Luttinger parameter one finds
\begin{equation}
    \frac{dK_\sigma}{dl} = - \frac{K_\sigma^2\tilde{y}^2_\sigma}{2} - (\tilde{J}^\pm_f)^2 - (\tilde{J}^z_b)^2
\end{equation}
in line with the fact that the flow is towards strong coupling, generating a spin gap by pinning $\phi_\sigma$ at the minimum of the cosine.
\bibliography{ref}
\end{document}